\documentclass[aps,pra,showpacs,groupedaddress,twocolumn,amsmath,amssymb]{revtex4-1}

\usepackage[american]{babel}
\usepackage{graphicx}
\usepackage{times}
\usepackage{color}

\begin{document}


\title{Nonclassicality Quasiprobability of Single-Photon Added Thermal States}

\author{T. Kiesel and W. Vogel}

\affiliation{Arbeitsgruppe Quantenoptik, Institut f\"ur Physik, Universit\"at  Rostock, D-18051 Rostock,
Germany}

\author{M. Bellini and A. Zavatta}
\affiliation{Istituto Nazionale di Ottica, INO-CNR, L.go E. Fermi, 6, I-50125, Florence, Italy}
\affiliation{LENS, Via Nello Carrara 1, 50019 Sesto Fiorentino, Florence, Italy}

\begin{abstract}
We report the experimental reconstruction of a nonclassicality quasiprobability for a single-photon added thermal state. This quantity has significant negativities, which is necessary and sufficient for the nonclassicality of the quantum state. Our method exhibits several advantages compared to the reconstruction of the $P$~function, since the nonclassicality filters used in this case can regularize the quasiprobabilities as well as their statistical uncertainties. A priori
assumptions about the quantum state are therefore not necessary. We also demonstrate that, in principle, our method
is not limited by small quantum efficiencies.
\end{abstract}


\maketitle

\section{Introduction.} The relation between classical optics and quantum optics is a fundamental topic in modern physics. The notion of nonclassicality has been introduced by Titulaer and Glauber as the impossibility to describe optical field correlations of a specific state of light in terms of classical electrodynamics \cite{pr-140B-676}. Therefore, nonclassical states are the prerequisite of quantum effects, which are of great interest in quantum optics and quantum information.

The definition of nonclassicality is based on the Glauber-Sudarshan phase-space representation of a quantum state $\hat\rho$,
\begin{equation}
 	\hat\rho = \int d^2\alpha P(\alpha)|\alpha\rangle\langle\alpha|,
\end{equation}
where $|\alpha\rangle$ denote the well-known coherent states~\cite{prl-10-277,pr131-2766}. In general, $P(\alpha)$ is a quasiprobability. If it has the properties of a classical probability distribution, the state $\hat \rho$ is a statistical mixture of coherent states, which are closely related to the classical behavior of the oscillator. Conversely, a state is referred to as nonclassical if the $P$~function shows some negativities. However, for many states, already for the single-photon state, this quantity is highly singular. Only in a few cases, when $P(\alpha)$ does not show singularities, it may be accessible from experimental data. Single-photon added thermal states (SPATS) belong to such a class of states~\cite{pra-46-485}, so that their $P$~function could be reconstructed for some parameters~\cite{pra-78-021804R}.

Due to the singularities of the $P$~function, several different nonclassicality criteria have been developed. Simple inequalities often set bounds for classical states, which are violated for certain nonclassical states; we only mention bounds on moments, such as (higher-order) squeezing~\cite{prl-54-323}, and probability distributions~\cite{pra-79-042105}. However, violation of these criteria is only sufficient, but not necessary for verifying nonclassicality. On the other hand, there are complete hierarchies of criteria, often based on matrices of moments \cite{pra-72-043808} or characteristic functions \cite{prl-89-283601,pra75-052106}. However, nobody is able to check such hierarchies completely, and practical application becomes involved for large matrices.

In \cite{pra-82-032107}, a novel approach for nonclassicality detection was developed, which was based on phase-space methods. It has been shown that for any nonclassical state, there exists a so-called nonclassicality quasiprobability which illustrates the nonclassical property by negativities. Moreover, it was shown that a family of nonclassicality distributions, parameterized by a real filter width $w$, enables one to decide whether a state is nonclassical or not. If a state is nonclassical, one can always find some finite filter width $w$, such that the nonclassicality is observable as a negativity of the corresponding nonclassicality quasiprobability. Our method is suitable for experimental application, since it incorporates correct handling of statistical uncertainties. Moreover, it does not require precognition about the state. 

In the present paper we demonstrate, by experimental application, the capability of the method. We examine single-photon added thermal states, whose $P$ function could be reconstructed for a sufficiently large mean thermal photon number~\cite{pra-78-021804R}. Here we overcome the problems occurring for 
arbitrary mean photon numbers, and therefore demonstrate the universality of the method of nonclassicality quasiprobabilities.

The paper is structures as follows: In section II, we briefly review the approximate reconstruction of the $P$ function of a SPATS and its limitations for relatively small mean photon numbers. In section III, we discuss the concept of nonclassicality quasiprobabilities and apply it to experimental data. Eventually in section IV, we consider the role of the quantum efficiency on the detection of nonclassicality. A summary and some conclusions are given in section~V.

\section{Approximate reconstruction of $P$~functions}


Let us briefly recall the reconstruction of a Glauber-Sudarshan representation as presented in \cite{pra-78-021804R}. The starting point of the discussion was the characteristic function $\Phi(\beta)$ of
a SPATS,
\begin{equation}
 \Phi(\beta) = \left[1-(1+\bar n)|\beta|^2\right] e^{-\bar n|\beta|^2}.
\end{equation}
By Fourier transform the resulting $P$~function is derived as
\begin{equation}
 P(\alpha) = \frac{1}{\pi \bar n^3}\left[(1+\bar n)|\alpha|^2-\bar n\right]e^{-|\alpha|^2/\bar n},
\end{equation}
which is a regular function.

Experimental data were used for a mean photon number of $\bar n = 1.11$ and a quantum efficiency of $\eta = 0.60$. The function $\Phi(\beta)$ was readily obtained from measured quadratures $\{x_j\}_{j=1}^N$ via
\begin{equation}
\label{eq:cf-samp}
 	\Phi(\beta) = \frac{e^{|\beta|^2/2}}{N}\sum_{j=1}^N e^{i|\beta| x_j}.
\end{equation}
It was directly sampled 
from $10^5$ data points. 
We observe that the experimentally obtained curve tends to zero within a fraction of its standard deviation,
\begin{equation}
\label{eq:cf-samp-var}
        \sigma^2\left\{\Phi(\beta)\right\}  = \frac{1}{N}\big[e^{|\beta|^2}-\left|\Phi(\beta)\right|^2\big].
\end{equation}
However, the latter grows exponentially with $|\beta|^2$. 
In order to calculate the Glauber-Sudarshan $P$~function via Fourier transform,
\begin{equation}
 P(\alpha) = \frac{1}{\pi^2}\int d^2\beta e^{\alpha\beta^*-\alpha^*\beta}\Phi(\beta),\label{eq:P:from:Phi}
\end{equation}
one has to regularize $\Phi(\beta)$. In our previous work, we simply cutoff $\Phi(\beta)$ for $|\beta| > |\beta_c|$. This is equivalent to the multiplication of $\Phi(\beta)$ with a rectangular filter, $\Omega_{\rm rect}(\beta)$, with $\Omega_{\rm rect}(\beta) = 1$ for $|\beta| < |\beta_c|$ and $\Omega_{\rm rect}(\beta) = 0$ elsewhere.
However, this method can only be applied in special cases. First, the state must be described by a well-behaved $P$~function, and its characteristic function has to decay sufficiently fast in order to justify the cutoff regularization. If this is not the case, one cannot perform the Fourier transform to obtain a $P$~function. Second, the systematic error has to be estimated by some assumptions on the behavior of the characteristic function for large $\beta$.
We used the theoretical expectation of the characteristic function with properly chosen parameters. For a completely unknown state, such a procedure becomes meaningless.
In the case of the SPATS, we obtained the $P$~function for mean thermal photon numbers of $\bar n \geq 1$, for details see~\cite{pra-78-021804R}. 

For smaller mean thermal photon numbers the reconstruction of the $P$~function faces severe limitations. The smaller $\bar n$, the broader the characteristic function becomes, and the larger is the statistical uncertainty at a reasonable cutoff parameter. Therefore, one cannot find a useful trade-off between the systematic and the statistical error. The former increase with lower $|\beta_c|$, and the statistical uncertainty is growing with larger $|\beta_c|$. For an example we refer the reader to the end of the next section. Under such circumstances, other nonclassicality criteria can be applied~\cite{pra75-052106}, which are sufficient but not necessary.

\section{Nonclassicality quasiprobabilities}

From a more general perspective to be used in the following, we may multiply the characteristic function by a filter function $\Omega(\beta)$,
\begin{equation}\label{eq:filter:phi}
	\Phi_\Omega(\beta)= \Phi(\beta)\Omega(\beta),
\end{equation}
for a general study of such a scenario cf.~\cite{Ag-Wo}. The special case discussed so far is contained in this approach for a rectangular filter. To obtain a quasiprobability -- including the full information on the quantum state under study -- by Fourier transform of $\Phi_\Omega(\beta)$, the filter must not have zeros anywhere in the complex plane.

In view of the radial symmetry of our quantum states, the Fourier transform of $\Phi_\Omega(\beta)$ is given by
\begin{equation}
\label{eq:P-reg:Hankel}
 P_\Omega(\alpha) = \frac{2}{\pi}\int_0^\infty b J_0(2b|\alpha|) \Phi_\Omega(b)d b.
\end{equation}
 This function, together with its variance,
\begin{widetext}
        \begin{equation}
        \sigma^2\left\{P_\Omega(\alpha)\right\} =
                \frac{1}{N}\left(\frac{4}{\pi^2}\iint_0^{\infty}\!\!\!\! b b' J_0(2b|\alpha|)J_0(2b'|\alpha|) \Phi(b-b')e^{b b'} \Omega (b)\Omega (b')\,d b'\, db - P_\Omega(\alpha)^2\right),
        \label{eq:var:P:Hankel}\,
        \end{equation}
\end{widetext}
can be readily calculated from the set of data. This expression for the variance of the quasiprobability is readily derived from the statistical sampling of the characteristic function according to Eq.~(\ref{eq:cf-samp}).

\subsection{Concept of nonclassicality quasiprobability}
Now we make use of the recently introduced concept of nonclassicality quasiprobabilities~\cite{pra-82-032107}.  
We introduce a so-called nonclassicality filter, $\Omega(\beta)\equiv\Omega_w(\beta)$,
with the following properties:
\begin{enumerate}
 	\item $\Omega_w(\beta)$ decays faster than $\exp(-|\beta|^2/2)$ for any filter width $w > 0$ in order to regularize the $P$ function and its statistical uncertainty for any quantum state.
	\item The Fourier transform of $\Omega_w(\beta)$ is non-negative, such that negativities in the nonclassicality quasiprobability are unambiguously caused by the negativity of the state's $P$~function.
	\item The parameter $w$ scales the filter $\Omega_w(\beta)$ such that for $w\to\infty$ the filter approaches one. Practically, we implement this by $\Omega_1(0) = 1,\ \Omega_w(\beta) = \Omega_1(\beta/w)$.
	\item The support of $\Omega_w(\beta)$ is the complex plane, such that Eq.~(\ref{eq:filter:phi}) is invertible for every $\beta$.
\end{enumerate}

The first requirement ensures that the integrals in Eq.~(\ref{eq:P-reg:Hankel}) and Eq.~(\ref{eq:var:P:Hankel}) are finite, when the filter $\Omega(\beta)$ is identified with the nonclassicality filter $\Omega_w(\beta)$. The second condition makes sure that the negativity of $P_\Omega(\alpha)$ unambiguously represents the nonclassicality of the state.
In contrast to this, a rectangular filter $\Omega_{\rm rect}(\beta)$ has a nonnegative Fourier transform.
Hence, for such a filter negativities of $P_\Omega(\alpha)$ must not be interpreted as
nonclassical effects without assumptions about the influence of the filter on the regularized $P$ function. In \cite{pra-78-021804R}, this made the estimation of systematic errors necessary.
The third condition can be used for maximizing the significance of the observed nonclassical effects. On the one hand, a larger filter width $w$ may increase the negativities in the regularized $P$~function, one the other hand, this will definitely increase the variance. We may tune $w$ in order to optimize the statistical significance $S$ of the negativities, defined  via
\begin{equation}
S=\frac{P_\Omega(\alpha)}{\sigma\{P_\Omega(\alpha)\}}.
\end{equation}
The fourth requirement is of fundamental importance: It ensures that the regularized function $P_\Omega(\alpha)$ still contains the full information on the quantum state. Such $P_\Omega$~functions,
which fulfill all the four conditions, we refer to as nonclassicality quasiprobabilities.

A proper nonclassicality filter can be constructed by the autocorrelation of a function $\omega(\beta)$,
\begin{equation}
 	\Omega_1(\beta) = \frac{1}{\mathcal N}\int d^2\beta' \omega(\beta') \omega(\beta+\beta'),
\end{equation}
with the normalization $\mathcal{N} = \int d^2\beta' |\omega(\beta')|^2$. The width parameter is introduced by $\Omega_w(\beta) = \Omega_1(\beta/w)$. The positivity of its Fourier transform is guaranteed by the properties of any autocorrelation function. Furthermore, if $\omega(\beta)$ decays sufficiently fast, as required by condition (1), $\Omega_w(\beta)$ does as well. Therefore, by choosing
\begin{equation}
	\omega(\beta) = e^{-|\beta|^4}
\end{equation}
and calculating the autocorrelation, we obtain a suitable representative of a nonclassicality filter.

\subsection{Experimental preparation of SPATS}

\begin{figure}[h]
\includegraphics*[width=\columnwidth]{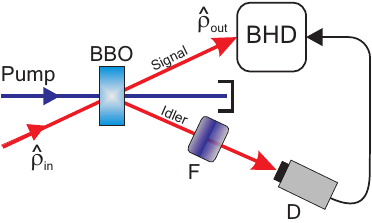}
\caption{Scheme for the conditional excitation of a thermal light state (denoted by $\hat \rho_{\rm in}$) by
a single photon. A click in the on-off detector D prepares the photon-added thermal state $\hat
\rho_{\rm out}$ and triggers its balanced homodyne
detection (BHD). \label{fig:setup}}
\end{figure}

The single-photon-added thermal states are generated in a conditional way by exploiting the parametric amplification at the single-photon level in a nonlinear type-I $\beta$-barium borate (BBO) crystal pumped by the second harmonic of a mode-locked picosecond Ti:sapphire laser (see Fig.~\ref{fig:setup}).

When no extra field is injected in the crystal, a pump photon can be converted into two spontaneously and simultaneously generated photons (named signal and idler) correlated in frequency and in momentum. The click of the on/off avalanche photodetector D, which is placed in the idler path after narrow spectral and spatial filters (F), is used to conditionally prepare a single photon in a well-defined spatiotemporal mode of the signal channel ~\cite{lvovsky01,pra04}.
On the other hand, if the parametric crystal is seeded with some light, described by the operator $\hat \rho_{\rm in}$, stimulated emission comes into play, and single-photon excitation of such a state, always converting it into a nonclassical one, is conditionally obtained when one photon is detected in the idler mode~\cite{science04,science07}. Field quadratures of the output signal state are then conditionally measured on a pulse-to-pulse basis using an ultrafast balanced homodyne detection scheme~\cite{josab02}.

Here we used a pseudo thermal source, obtained by inserting a rotating ground glass disk in a
portion of the laser beam (see~\cite{pra75-052106,arecchi65,parigi09}), for injecting the parametric amplifier and producing SPATS.

\subsection{Experimental nonclassicality quasiprobabilities}

\begin{figure}
 	\includegraphics[width=\columnwidth]{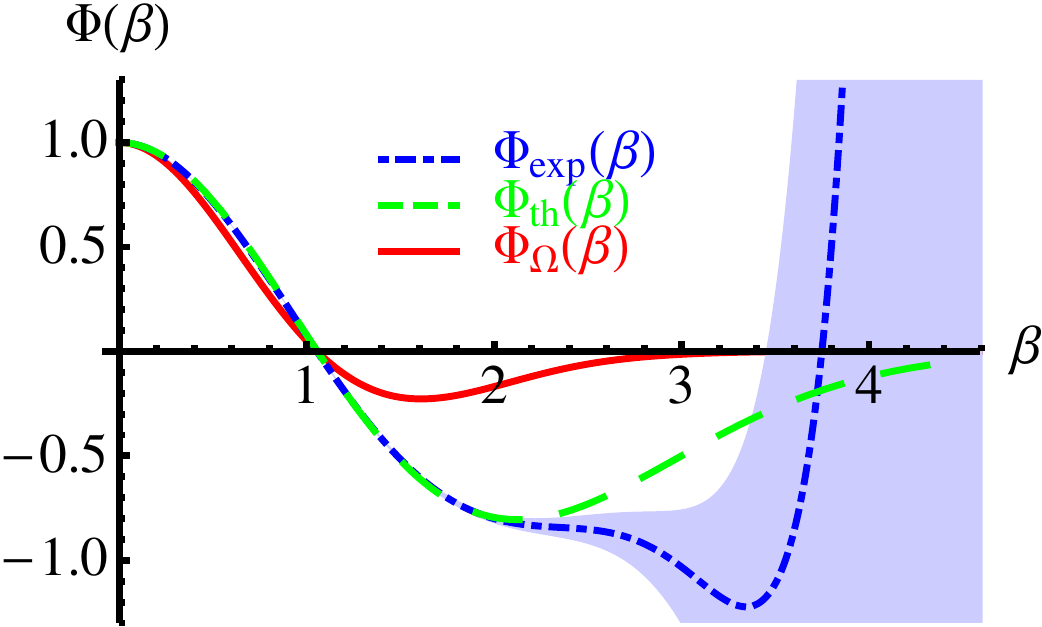}
	\caption{Characteristic functions of a SPATS with $\bar n = 0.49$ and $\eta=0.62$: the experimental result $\Phi_{\rm exp}(\beta)$,
the theoretical expectation $\Phi_{\rm th}(\beta)$, and the result $\Phi_{\Omega}(\beta)$ of filtering for the filter width $w = 1.4$.
The shaded area corresponds to one standard deviation.} 
	\label{fig:phi:0.49}
\end{figure}

To illustrate the power of nonclassicality quasiprobabilities, let us consider a SPATS with $\bar n = 0.49$ and $\eta = 0.62$. In Fig.~\ref{fig:phi:0.49}, we show the experimentally reconstructed characteristic function $\Phi_{\rm exp}(\beta)$. Obviously, the Fourier transform of this quantity does not exist as a regular function, since $\Phi_{\rm exp}(\beta)$ does not approach zero for large $\beta$ as its theoretical expectation $\Phi_{\rm th}(\beta)$ does. This is due to the fact that the uncertainty grows exponentially. Although both the experimental result and the theoretical expectation agree within two standard deviations, the former may even diverge within the divergent noise level. In contrast to the results in \cite{pra-78-021804R}, it is not obvious just from experimental data, that the characteristic function tends to zero for large $\beta$. Moreover, it is not possible to find a reasonable compromise between a low systematic error (requiring a large cutoff parameter $|\beta_c|$), and a statistical uncertainty being sufficiently small to obtain significant negativities in the filtered $P$ function.

\begin{figure}
 	\includegraphics[width=0.9\columnwidth]{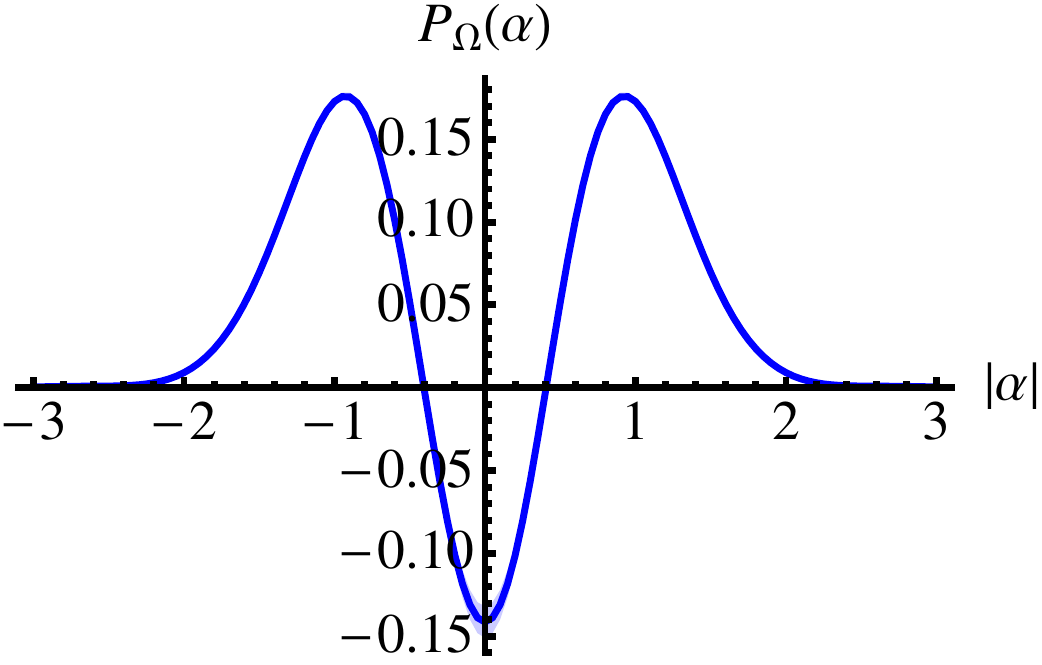}
	\caption{Nonclassicality quasiprobability of a SPATS for the same parameters as in Fig.~\ref{fig:phi:0.49}. The blue shaded area corresponds to one standard deviation.}
	\label{fig:P:0.49}
\end{figure}

The application of the nonclassicality filter with a width $w=1.4$ leads to an integrable characteristic function $\Phi_{\Omega}(\beta)$. We emphasize that the shown curve is obtained from the experimental data. We also calculated its standard deviation, which is included in the line thickness. Therefore, this function is suited for deriving the corresponding nonclassicality quasiprobability by Fourier transform. Figure~\ref{fig:P:0.49} shows the result. We observe a distinct negativity at the origin of phase space, with a significance of $15$ standard deviations. By the definition of nonclassicality quasiprobabilities, this negativity is solely due to the nonclassicality of the state.

We may also reconsider the data of the SPATS with $\bar n=1.11$ by our filtering procedure. We obtain a nonclassicality quasiprobability $P_\Omega(\beta)$ which looks similar to the one in Fig.~\ref{fig:P:0.49}. By optimizing the filter width to $w = 1.3$, we get a maximum significance of $7.6$ standard deviations, which exceeds that for the rectangular filter.
More importantly, the estimation of a systematic error -- which was previously needed for the assessment of nonclassicality and had been 
based on a-priori theoretical assumptions -- now becomes superfluous.

\subsection{Limits of rectangular filtering}

\begin{figure}[h]
    \includegraphics[width=\columnwidth]{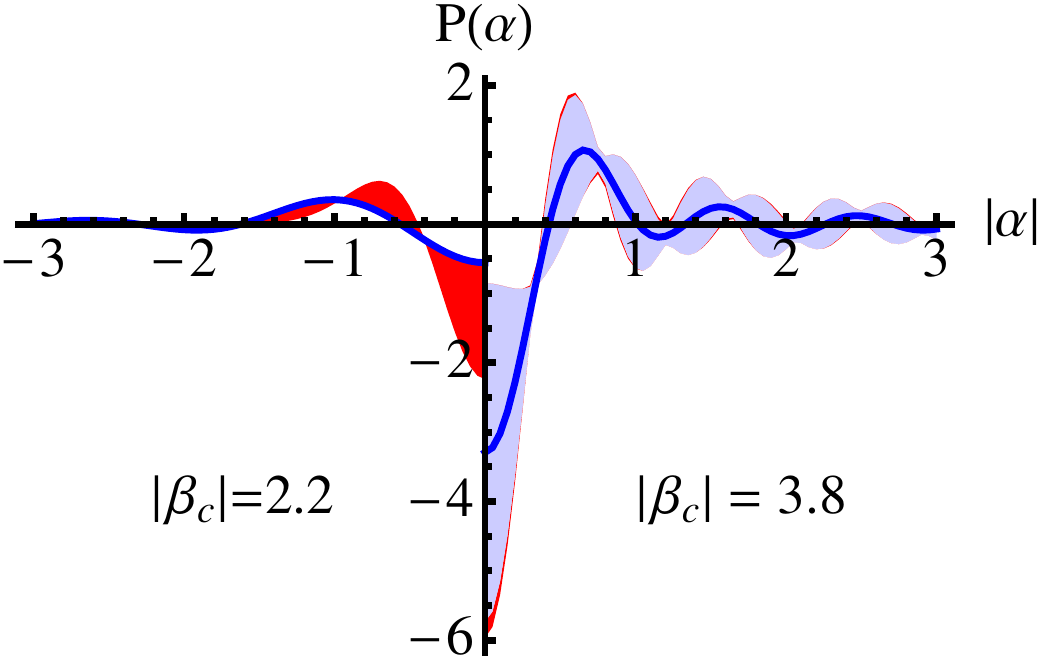}
    \caption{(color online) $P$ functions regularized with a rectangular filter. The cutoff parameter at the left side of the figure is rather small, leading to large systematic errors (red shaded). For a larger cutoff parameter, the statistical uncertainty (blue shaded) becomes dominant, see the right side of the figure.}
    \label{fig:Rectfilter:TwoSettings}
\end{figure}

For comparison, let us try to regularize the state with $\bar n = 0.49$ and $\eta = 0.62$, whose characteristic function is shown in Fig.~\ref{fig:phi:0.49}, with a rectangular filter. Since this is not a nonclassicality filter of the type defined in Sec.~III~A, we have to consider the systematic error, which comes from the cutoff. From Fig.~\ref{fig:phi:0.49}, two possible cutoff parameter may be reasonable: On the one hand, we may choose $|\beta_c| = 2.2$, since from this point the deviations of the theoretical and experimental characteristic functions $\Phi_{\rm th}(\beta)$ and $\Phi_{\rm exp}(\beta)$ grow strongly. The corresponding regularized $P$ function with its standard deviation (shaded blue area) and systematic error (dark red area) is shown at the left side of Fig.~\ref{fig:Rectfilter:TwoSettings}. We clearly see that the statistical uncertainty is negligible, while the systematic error is partly even larger than of the $P$ function. Therefore, a-priori assumptions about the state, which are the basis for the estimation of the systematic error, are crucial. Moreover, the systematic error is larger than the size of the negativity.

On the other hand, we may set $|\beta_c| = 3.8$, where the experimental characteristic function $\Phi_{\rm exp}(\beta)$ is close to zero. The resulting $P$ function is shown at the right side of Fig.~\ref{fig:Rectfilter:TwoSettings}.  In this case, the systematic error is small, while the statistical uncertainty is dominating. The significance of the negativity of the filtered $P$ function is less than $1.4$ standard deviations.

Therefore, we are not able to obtain statistically significant negativities by applying a rectangular filter, if we simultaneously  try to achieve low  systematic errors due to the regularization. The reason is that the characteristic function of a state with such a small mean photon number approaches zero only for larger $\beta$, where the standard deviation is already extremely large. This problem can only be overcome by using suitable nonclassicality filter, which do not require the consideration of systematic errors in order to verify nonclassicality.


\section{Influence of the quantum efficiency}

Let us assume that the characteristic function $\Phi(\beta)$ describes a quantum state, but we measure this state with a non-unit quantum efficiency $\eta$. The characteristic function of the measured state is given by
\begin{equation}
 	\Phi(\beta;\eta) = \Phi(\sqrt{\eta}\beta).
\end{equation}
Since the corresponding $P$ function is the Fourier transform, this rescaling of $\beta$ leads to a rescaling of the argument of the $P$ function,
\begin{equation}
 	P (\alpha;\eta)= \frac{1}{\eta}P\left(\frac{\alpha}{\sqrt{\eta}}\right).
\end{equation}
Therefore, a non-unit quantum efficiency does not degrade nonclassicality, and removal of losses from experimental data only rescales the $P$ function, but does not affect its course and uncertainty. In contrast, this is not the case for the Wigner function whose negativities more and more disappear with decreasing $\eta$-values.

How is the situation for nonclassicality quasiprobabilities? For answering this question, we note that the filtered characteristic function of a state, which suffered losses, reads as
\begin{equation}
	\Phi_{\Omega}(\beta;\eta) = \Phi(\beta;\eta)\Omega_w(\beta) = \Phi(\sqrt{\eta}\beta)\Omega_1(\beta/w).
\end{equation}
We observe that the characteristic function of the lossy state, $\Phi_{\Omega}(\beta;\eta)$, and the one of the ideal state, $\Phi_{\Omega}(\beta;1)$ are not connected by simple rescaling. However, they can be easily connected if one allows to rescale the filter width as well:
\begin{equation}
	\Phi_{\Omega}(\beta; \eta) =  \Phi(\sqrt{\eta}\beta)\Omega_1((\sqrt{\eta}\beta)/(\sqrt{\eta}w)).
\end{equation}
Hence, the characteristic function of a lossy state with some width $w$, which is on the left side of the equation, is given by the characteristic function of the ideal state with filter width $\sqrt{\eta}w$ by additionally rescaling the argument $\beta\to\sqrt{\eta}\beta$. Removing the losses by post-processing and simultaneously adapting the width of the nonclassicality filter leads simply to scaled results with no better significance. Therefore, the removal of losses does not uncover nonclassical effects which do not already appear in the nonclassicality quasiprobabilities of the lossy state.

Finally, we note that our method may visualize nonclassicality even for rather small quantum efficiencies. Simulations show that we find for the SPATS with $\bar n =0.49$, for our sample of $10^5$ data points, negativities in the nonclassicality quasiprobability with a significance of at least three standard deviations, if $\eta \geq 0.4$. This $\eta$-value is only limited from below by the size of the sample. In contrast,  for $\eta \leq 0.5$ the Wigner function is always nonnegative. Therefore, the negativities of the nonclassicality quasiprobabilities are more sensitive to nonclassical effects than negativities of the Wigner function.

\section{Summary and Conclusions}

We have applied the concept of nonclassicality quasiprobabilities to experimental data of single-photon added thermal states. Even though the Glauber-Sudarshan $P$~function of these states is regular in general, its approximate reconstruction is feasible only for a certain parameter range. Moreover, it requires to make use of some  precognition on the state under study. Our quasiprobability approach does not require such a precognition, it works for any quantum state -- even when the $P$~function is strongly singular -- and it also suppresses the experimental sampling noise.

We have shown that the nonclassicality filters needed in our procedure, suppress the exponential growth of experimentally determined characteristic functions, which yields integrable functions. Hence, Fourier transform delivers nonclassicality quasiprobabilities with finite statistical uncertainties. By optimization of the filter width, significant negativities in the quasiprobabilities are found for nonclassical states. For our chosen example, an approximate reconstruction of the $P$~function was shown to be no longer useful. 

With the accessible set of $10^5$ data points we could demonstrate negativities in the experimentally determined nonclassicality quasiprobability with a significance of about $15$ standard deviations. This result is solely limited by the statistical uncertainties caused by the finite size of the available set of experimental data, it could be further improved by extending this set.
We have also considered the role of imperfect detection. In fact, the detection efficiency can be completely eliminated by a proper rescaling of arguments in our functions. Thus, even with a rather small efficiency one can identify all nonclassical effects, provided the sampling noise is sufficiently small.


\paragraph*{Acknowledgements.}
This work was supported by the Deutsche Forschungsgemeinschaft through SFB 652, and partially supported by Ente Cassa di Risparmio di Firenze and Regione Toscana under project CTOTUS.

\end{document}